# The role of Ca in superconducting and magnetic properties of $Y_{1-x}Ca_xBa_2Cu_3O_{7-\delta}$

## (x = 0.0-0.30)

N.P Liyanawaduge[1,2,3], Anuj Kumar[2], Shiva Kumar[2], B.S.B Karunarathne[3] and V.P. S. Awana[2*]

[1]Industrial Technology Institute, P.O Box 363, Baudhaloka Mawatha, Colombo-07, Sri Lanka

[2]Quantum Phenomena and Application Division, National Physical Laboratory (*CSIR*)

New Delhi-110012, India

[3]Department of Physics, University of Peradeniya, Peradeniya (20400), Sri Lanka

## Abstract

We report the structural, superconducting, magnetic and granular properties of $Y_{1-x}Ca_xBa_2Cu_3O_{7-\delta}$; x = 0.0, 0.1, 0.2 and 0.3. Rietveld fitted X-ray diffraction data confirms the single phase formation for all the samples. The orthorhombicity of parent compound (x = 0.0) decreases and becomes closer to the tetragonal structure for higher Ca concentration. The superconducting transition temperature ($T_c$) decreases with increasing Ca content due to both over-doping and ensuing disorder in the superconducting $CuO_2$ planes with Ca doping. DC susceptibility measurements reveal reduction of Meissner fraction with Ca doping, suggesting the flux pining effect. The AC susceptibility measurements reveal the enhancement of grain coupling with increasing Ca content in the system. The average grain size is found to increase with Ca doping. The Scanning Electron Microscopy (*SEM*) observations indicate better grain connectivity in terms of narrow grain boundaries for Ca doped samples. It is concluded that limited Ca doping enhances the superconducting performance of $YBa_2Cu_3O_{7-\delta}$ system.

## 1. Introduction

The carrier density in $YBa_2Cu_3O_{7-\delta}$ (YBCO) can be enhanced either by increasing oxygen stoichiometry $(7-\delta)$ [1] or by on-site ionic substitutions [2,3]. The substitution of lower valance state cations at the $Y^{3+}$ site in YBCO system is of great interest because it directly affects the carrier concentration in adjacent superconducting $CuO_2$ planes. Also, it is expected that by such substitution, the charge distribution in the structure and consequently the concentration of charge carriers could be varied mostly at wish [2]. Calcium $(Ca^{2+})$ is one of the suitable candidates, which preferentially substitutes for $Y^{3+}$ because its ionic radius is closer to the Yttrium. This leads to many interesting physical property changes in YBCO such as; critical temperature $(T_c)$, normal state conductivity $(\rho_n)$, thermoelectric power (S) and lattice parameters etc.

$Ca^{2+}$ substitution for $Y^{3+}$ increases the overall hole concentration [2, 4]. Fisher *et al*. [4] calculated the Hall number at different oxygen contents and concluded that the hole doping by calcium is not uniform and depend on oxygen content of the system. Hole doping is very low for highly oxygenated samples, while it is quite significant for oxygen deficient samples. The hole content in oxygenated Ca doped YBCO samples is very close to that of highly oxygenated un-doped samples. It is seen that in low $\delta$ regime, the compensating effect of calcium for hole filling is marginal [2, 4]. This hole doping effect, first takes place in $CuO_2$ planes and after that causes oxygen removal from $CuO_{1-\delta}$ chains. The incorporation of calcium into the lattice is accomplished by large oxygen vacancies [2, 5, 6]. The oxygen removal counteracts the hole doping behavior of calcium to the YBCO system followed by structural transformation from orthorhombic to tetragonal structure [2, 5, 6].

The overall hole density depend on the competition of hole doping behavior of Ca and hole vacancies being created due to the removal of oxygen [2, 4, 6]. As the number of holes increases, the in-plane Cu(2)-O(2) bond length get shortened and lattice parameter $a$ decreases. This also causes increase in Cu(2)-O(2)-Cu(2) angle and thus CuO$_2$ planes become flatter [6]. Further addition of Ca, increases the hole density of the system and tends to removal of oxygen from the CuO$_{1-\delta}$ chains to compensate the additional effect of Ca. This results in increase of the lattice parameter $a$, and decrease in the orthorhombocity of the structure [6]. Whereas the Ca doping into the oxygen rich YBCO shows different behavior and it creates oxygen vacancies, increase in in-plane Cu(2)-O(2) bond distance, $a$ parameter and the Cu(2)-O(2)-Cu(2) angle [7, 8].

The superconducting transition temperature (T$_c$) variation upon the hole doping occurs in two different ways as per oxygen content of the system. For oxygen deficient or partially oxygenated structure, the T$_c$ increases with increasing Ca concentration by enhancing the overall hole density [2, 6, 9]. In contrast, in oxygen rich samples Ca substitution causes reduction of T$_c$ [2, 5, 7-8, 10-11]. This T$_c$ depression is unlikely to be due to over-doping alone. Pure YBCO is slightly over-doped near $\delta = 0$ [2, 12] therefore it shows reduction of T$_c$ in highly oxygenated regime. The detrimental behavior of T$_c$ in Ca substituted samples is due to the over-doping effect with disorder being created in superconducting CuO$_2$ planes [9-11]. In addition, Shakeripour *et al*. [8] concluded that the reduction of T$_c$ with increasing Ca in oxygen rich system is due to the flattening of the CuO$_2$ planes. Therefore the induction of mobile carriers, overall oxygen content, oxygen disorder and structural changes are the factors which are responsible for T$_c$ behavior in Y$_{1-x}$Ca$_x$BCO system.

In this work we synthesized $Y_{1-x}Ca_xBa_2Cu_3O_{7-\delta}$ (x = 0.0-0.03) samples closer to the oxygen rich regime and investigated the structural, superconducting, magnetic and granular properties. We report XRD analysis, resistivity behavior ($\rho$-T), AC/DC linear susceptibility and SEM of these samples. It is concluded that grains coupling improves for moderately Ca doped $YBa_2Cu_3O_{7-\delta}$ samples.

## 2. Experimental

Samples are synthesized through conventional solid state reaction route. High purity powders of $Y_2O_3$, $CaCO_3$, $BaCO_3$ and CuO and (99.99%) with exact stoichiometric ratio are mixed. After initial grinding, calcination is done at $850^oC$ in air for 36 h and cooled to the room temperature. Subsequent calcinations are done at $870^oC$, $890^oC$ and $910^oC$ temperatures with intermediate grinding. In each calcinations cycle cooling is done slowly and samples are re-ground well before the next cycle. After final calcination the samples are pressed into rectangular pellet form and sintered at $915^oC$ for 48 h in air. Finally the samples are annealed with flowing oxygen at $900^oC$ for 12 h, $600^oC$ for 12 h and $450^oC$ for 12 h and subsequently cooled slowly. All the samples are characterized by the X-ray powder diffraction technique using Rigaku X-ray diffractometer (Cu-K$_\alpha$ line). Rietveld analysis of all samples is performed using Fullprof program. Resistivity measurements of all samples are performed with standard four probe method. The AC/DC susceptibility is measured by Physical Properties Measurement System (Quantum Design-USA PPMS-14Tesla). The SEM images of the samples are taken using *ZEISS* EVO MA-10 Scanning Electron Microscope.

### 3. Results and Discussion

The Rietveld fitted *XRD* patterns for studied $Y_{1-x}Ca_xBa_2Cu_3O_{7-\delta}$; x = 0.0, 0.1, 0.2 and 0.3 are shown in Fig. 1. It is confirmed that all the samples crystallize in single phase formation with orthorhombic *pmmm* space group. The lattice parameters and other fitted parameters are given in Table 1. It can be seen from the *XRD* patterns and fitted lattice parameters, that with increase in Ca concentration, the orthorhombicity of the structure decreases, which is in agreement with early studies [2, 5, 8]. Cava *et al.* [1] measured the lattice parameters with different oxygen stoichiometry for pure YBCO. The lattice parameters, calculated through Rietveld analysis for YBCO in this study, are close to the cell parameters of the sample with the highest oxygen stoichiometry of close to 7.0, being reported in ref. [1]. Therefore it is reasonable to assume that samples being investigated in current study are close to the oxygen rich regime. We calculated the lattice parameters of $Y_{1-x}Ca_xBa_2Cu_3O_{7-\delta}$; x = 0.0, 0.1, 0.2 and 0.3 by Reitveld fitting of the observed XRD patterns. It is found that though *a*-parameter increases, the *b*-parameter decreases with Ca doping (see Table 1), which is in agreement with earlier reports on fully oxygenated $Y_{1-x}Ca_xBa_2Cu_3O_{7-\delta}$ system [2, 5, 7-8, 10-11]. It is known that extra hole carriers are introduced to the $CuO_2$ planes with increasing Ca concentration [2-4]. This is counteracted by the oxygen vacancies in $CuO_{1-\delta}$ chains thus created. The same results in an increase of *a*-parameter and decrease of *b*-parameter. The *c* lattice parameter is slightly increasing (Table 1) with Ca concentration, again in agreement with earlier reports [2, 5, 7-8, 10-11]. Our XRD results (Fig.1) and lattice parameters variation (Table 1) conform that our studied samples are well oxygenated and the substitution of Ca at Y site in $Y_{1-x}Ca_xBa_2Cu_3O_{7-\delta}$ is successful.

The resistivity plot of all the samples shows that the $T_c$ value decreases with increasing Ca content [Fig. 2]. The simultaneous variation of lattice parameters and $T_c$ with Ca addition confirms that Ca substitutes at Y site. The samples up to x=0.2 show sharp transition but x=0.3 concentration exhibits relatively broader transition with two steps. This discrepancy is due to partial decomposition taking place in the sample with x=0.3 [2]. Sintering temperature needed for good inter-grain connectivity causes partial decomposition in highly Ca doped samples [2]. Since these samples are already oxygen rich, further increase of Ca leads over-doping and also induces disorder in the superconducting $CuO_2$ planes [11]. A critical oxygen content, $x_c$, is reported [5] such that disorder of the oxygen has no effect when $x < x_c$ but it creates mobile hole traps when $x > x_c$. The value of $x_c$ gets lower with increase of Ca content. Therefore reduction in $T_c$ with increasing Ca content is not only due to over-doping but also due to the oxygen disorder effect. Also the flattening of $CuO_2$ planes causes decrease in the internal crystallographic pressure of the system and thus decreases the $T_c$ [8-13]. The normal state resistivity decreases up to x = 0.20 due to the hole doping effect of Ca and subsequently increases in x = 0.30 due to the possible decomposition.

Fig. 3 presents the variation of DC volume susceptibility curves in both Field Cooled (FC) and Zero Field Cooled (ZFC) conditions. The variation of diamagnetic transition temperature is in agreement to the $\rho$-T plots. The saturated ZFC diamagnetic magnetization, which is measure of diamagnetic shielding current increases systematically up to x = 0.20. Also, the separation of FC and ZFC signals increases monotonically until x = 0.20 sample and nearly saturates for x = 0.30. The increase in separation of FC and ZFC signals reflects reduction of Meissner fraction (ratio of field cooled to zero fielded cooled magnetization), which is an indication of stronger flux trapping process. These observations claim a pinning like behavior is developing with Ca doping. This is evident from decreased FC signal with increase in Ca content. Therefore, an enhancement of critical current density, $j_c$ can be expected with increasing Ca content. It seems the effective pinning is taking place till x = 0.20. In case of x = 0.30 possible decomposition being reminiscent from resistivity results hinders the effective pinning process. It is worth mentioning here that the increased separation of FC and ZFC magnetization along with decreasing FC signal though clearly suggest towards strong pinning, the exact measurements of critical current density, $j_c$ need to be carried out to ensure the effective pinning. The high field magnetization (MH) and magneto-transport R(T)H of these samples could unearth the moot question of effective pinning, which at present we leave for future study on the same samples.

AC linear susceptibility of all these samples measured at different AC field amplitudes with zero bias dc field is given in Fig. 4. It is well known that the real part of susceptibility consists of two transitions which correspond to the flux removal from intra-grain and inter-grain regimes. In accord, imaginary part contains two peaks which represent energy dissipation and ac losses due to the flux motion in intra-grain and inter-grain regions [14-18]. The amplitude of high temperature peak in the imaginary part corresponding to the individual grains is a measure of the grain size [17, 18]. In Fig. 4 (a) it is noticeable that the position (temperature) of high temperature peak is almost insensitive to the applied field while the position of low temperature peak due to inter-grain couplings is highly sensitive to the field amplitude. At higher fields the field can penetrate the sample deeper than in lower fields. Therefore both real and imaginary signal enhances with increasing field amplitude. The inter-grain peak in Fig.4 (a) lies very close to the intra-grain peak at lower fields but it shifts to the low temperatures at higher fields. In fact at lower fields it is difficult to distinguish intra-grain and inter granular regions to applied field due to persisting strong inter-grain coupling. But at moderate field amplitudes it can separate inter-grain and intra-grain regimes separately and show well distinguished inter-grain and intra-grain signals [14-18]. Interestingly, Ca doped samples [Fig. 4 (b), Fig. 4 (c) and Fig. 4 (d)] behave in different way than in the pristine sample and show very little shift of inter-grain peak towards lower temperature. This behavior enhances with increasing Ca content. This is due to the strong inter-grain couplings induced to the system with the doping of Ca. The fields, strong enough to separate inter-grain and intra-grain interactions in pure sample, are now not enough to separate these interactions in doped samples.

In the pristine sample the intra-granular signal enhances with increasing field and the best enhanced intra-grain signal is about 15 Oe (higher field). But in sample with x = 0.10, intra-granular signal is not visible for any applied field. This can be due to the introduction of strong inter-granular interactions with Ca doping even though slight increase in the grain size is observed in SEM image (to be discussed). This may suppress the granular signal in sample with x = 0.10. The samples with x = 0.20 and 0.30 again show a granular property but at relatively lower fields (the most enhanced intra-grain signals are seen around 9 Oe and 3 Oe respectively). This can be due to increase of grain size with Ca addition as revealed by SEM images. The pristine sample possesses smaller grains and resultant more grain boundaries (see Fig.6) hence it needs higher fields (15 Oe) to penetrate sample deeper to give significance intra-granular peak. On the other hand the Ca doped samples possess relatively bigger grains with less resultant grain boundaries and thus the field required to give a significant intra-grain peak is decreasing with x = 0.20 and 0.30 the same is 9 Oe and 3 Oe respectively. But the intra-grain signal is suppressed quickly, as inter-granular signal enhances more rapidly than the intra-grain signal (increasing grain size) due to the persisting strong grain couplings in doped samples.

Fig. 5a and 5b depict the real and imaginary components respectively of AC susceptibility measured at 9 Oe and 333 Hz AC field. It can be seen here also that the same pattern of variation of diamagnetic transition temperatures with Ca concentrations as in DC magnetization curves. In Fig. 5a, only pristine sample show well separated two step transitions with positive slope at onset, while doped samples show sharp transitions and negative slope at onset. In doped samples the second transition shifts to higher temperatures and gets overlapped with first transition due to persisting strong grain couplings [14-18]. In accord the two peaks in Fig. 5b behave in same fashion as two transitions in Fig. 5a. Once again it can be noted in Fig. 5b, that the amplitude of granular peak has enhanced in doped samples than in pure sample. These observations reflect that the Ca doped samples shows enhanced grain couplings which can increase critical current density in doped samples.

The observations and predictions from AC/DC magnetization data can be emphasized in light of SEM images of these samples given in Fig. 6. It is noticeable that doping of Ca even in small amount causes changes in grain morphology dramatically. The round shaped well separated grains are grown in pure sample, while capsule shaped attached grains are developed in doped samples. We have calculated the average grain size using the standardized linear intercept (LI) method and encircled the average grain in corresponding SEM image. It is clear that the average grain size increases with Ca doping. In the pure sample the grains can be identified separately and grain boundaries are well defined. Where as in doped samples (x=0.2 and 0.3) some sort of aggregation of grains can be seen and somewhat difficult to distinguish the grains within a single cluster. Thus improved grain connectivity between grains is observed. It was seen that two steps transition exists in $\rho$-T plot and a relatively broader transition in the AC/DC magnetization curves of sample with x = 0.3. This behavior may arise due to formation of clusters of grains like nature along with decomposition. Further, very fine impurity particles grown on grains can be seen for x = 0.3 [Fig. 6d]. Even though lower fields able to give arise two signals for inter and intra interactions, higher fields see the cluster as a single unit [Fig. 4d]. This is possible as inter-grain signal pronounce more rapidly than intra-grain signal at higher fields in higher Ca doped samples.

As mentioned earlier, the magnetization measurements (FC-ZFC separation) indicate towards flux pining, which could enhance the critical current density $J_c$ of the system. This is possible due to the increment of effective number of charge carriers in the matrix with increasing Ca content and hence increasing the effective current carrying through cross section [22, 20-22]. This is facilitated by inducing extra holes at grain boundary regions via controlling oxygen deficiency [20].  In doped samples, Ca not only goes to the grains but also to grain boundaries [23] due to its high segregation ability [24]. The Ca mainly replaces Y in grain boundary regions as well as within the grains. But in order to relieve persisting compressive or tensile stress at the grain boundaries, the Ca also replaces Cu and Ba sites respectively and reduce the formation of O vacancies [20].

Summarily, our magnetization and AC susceptibility results indicate that Ca doping into the YBCO matrix develops a flux pining mechanism and hence effectively may increase the $J_c$ of the system. This is physically strengthened by SEM results. The high field magnetization and magneto-transport studies are yet desired on these samples to ascertain the envisaged increased $J_c$ as a future study.

## 5. Acknowledgement

This work is supported by *RTFDCS* fellowship programme conducted by Center for International Cooperation in Science (*CICS*). Authors are thankful to Prof. R.C Budhani, Director *NPL* and to Dr. (Mrs.) Ganga Radhakrishnan, Director *CICS* for their encouragement. Authors Anuj Kumar and Shiva Kumar would like to acknowledge *CSIR*, *India* for providing fellowship.

**Figure Caption and Table**

**Fig. 1** Rietveld fitted XRD pattern of $Y_{1-x}Ca_xBa_2Cu_3O_{7-\delta}$, x: 0.0, 0.1, 0.2, 0.3, oxygen annealed samples.

**Fig. 2** Temperature dependency of resistivity of $Y_{1-x}Ca_xBa_2Cu_3O_{7-\delta}$, x: 0.0, 0.1, 0.2, 0.3, oxygen annealed samples.

**Fig. 3** Variation of dimensionless DC volume susceptibility with temperature of $Y_{1-x}Ca_xBa_2Cu_3O_{7-\delta}$ samples annealed in oxygen environment.

**Fig. 4** Real part and imaginary part of AC magnetization of $Y_{1-x}Ca_xBa_2Cu_3O_{7-\delta}$, (a) x = 0.0, (b) x = 0.1, (c) x = 0.2, (d) x = 0.3 taken at different ac field amplitudes with 333Hz constant frequency.

**Fig. 5** (a) Real part (b) Imaginary part of AC susceptibility of $Y_{1-x}Ca_xBa_2Cu_3O_{7-\delta}$ performed at 9 Oe and 333Hz ac field.

**Fig. 6** SEM images of $Y_{1-x}Ca_xBa_2Cu_3O_{7-\delta}$; (a) x = 0.0, (b) x = 0.1, (c) x = 0.2, (d) x = 0.3 at magnification 10 KX. The average grain is encircled in each image

**Table 1** Rietveld refined parameters of $Y_{1-x}Ca_xBa_2Cu_3O_{7-\delta}$, oxygen annealed samples.

| $Y_{1-x}Ca_xBa_2Cu_3O_{7-\delta}$ | a (A°) | b (A°) | c (A°) | $R_p$ | $R_{wp}$ | $\chi^2$ |
|---|---|---|---|---|---|---|
| x = 0.0 | 3.822 | 3.887 | 11.686 | 5.08 | 6.57 | 3.40 |
| x = 0.1 | 3.826 | 3.881 | 11.694 | 4.91 | 6.28 | 2.89 |
| x = 0.2 | 3.831 | 3.877 | 11.703 | 4.80 | 6.19 | 2.63 |
| x = 0.3 | 3.836 | 3.873 | 11.711 | 5.10 | 6.60 | 3.65 |

**Fig.1**

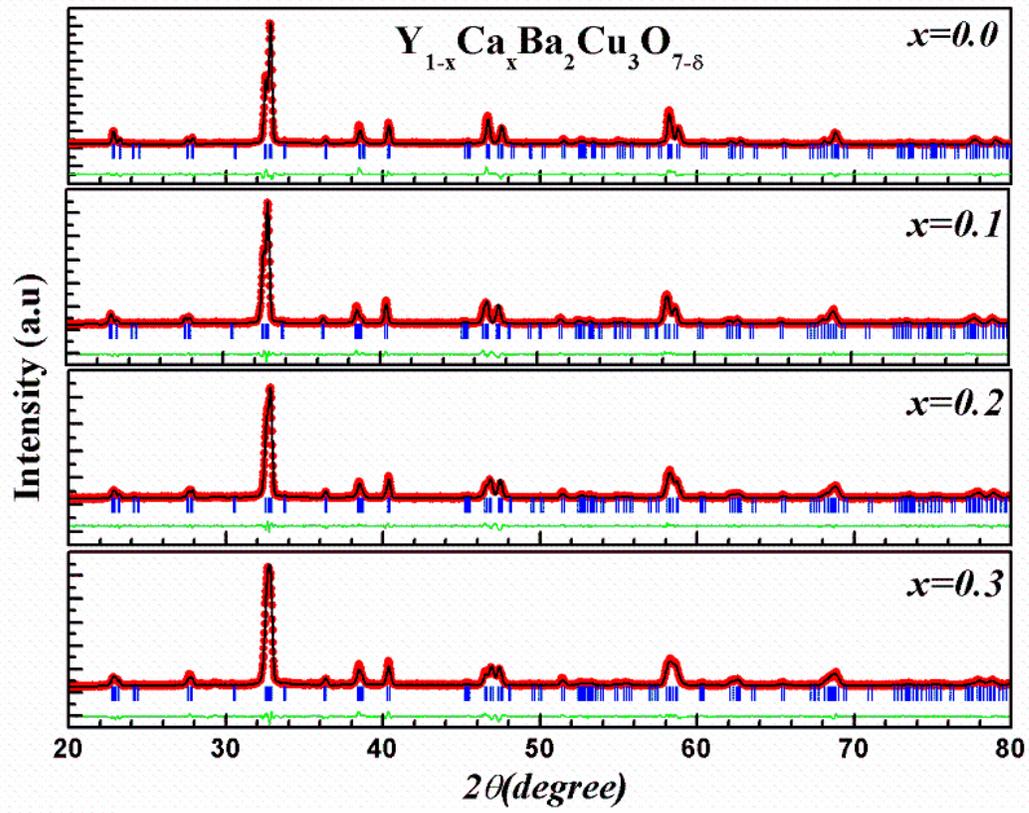

**Fig.2**

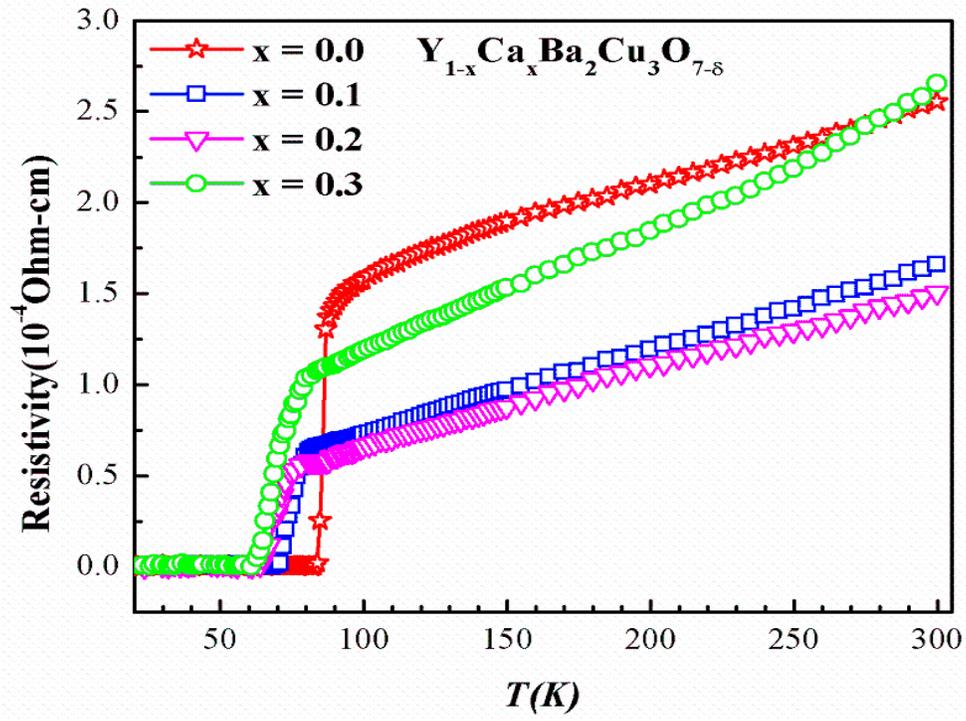

**Fig.3**

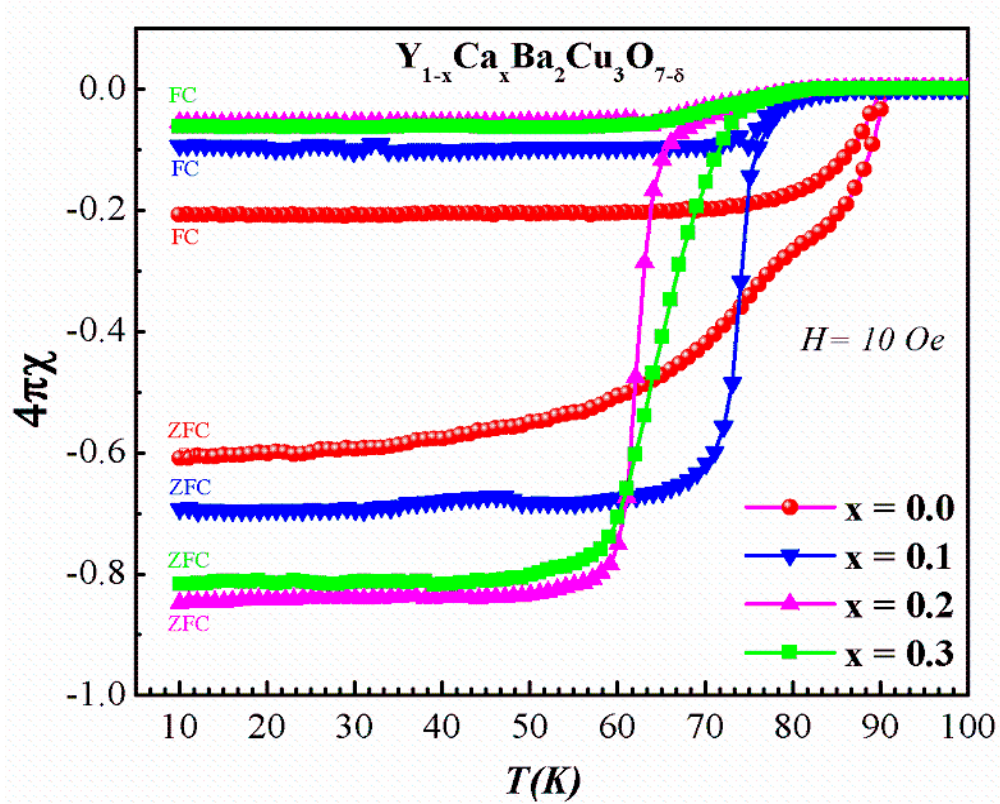

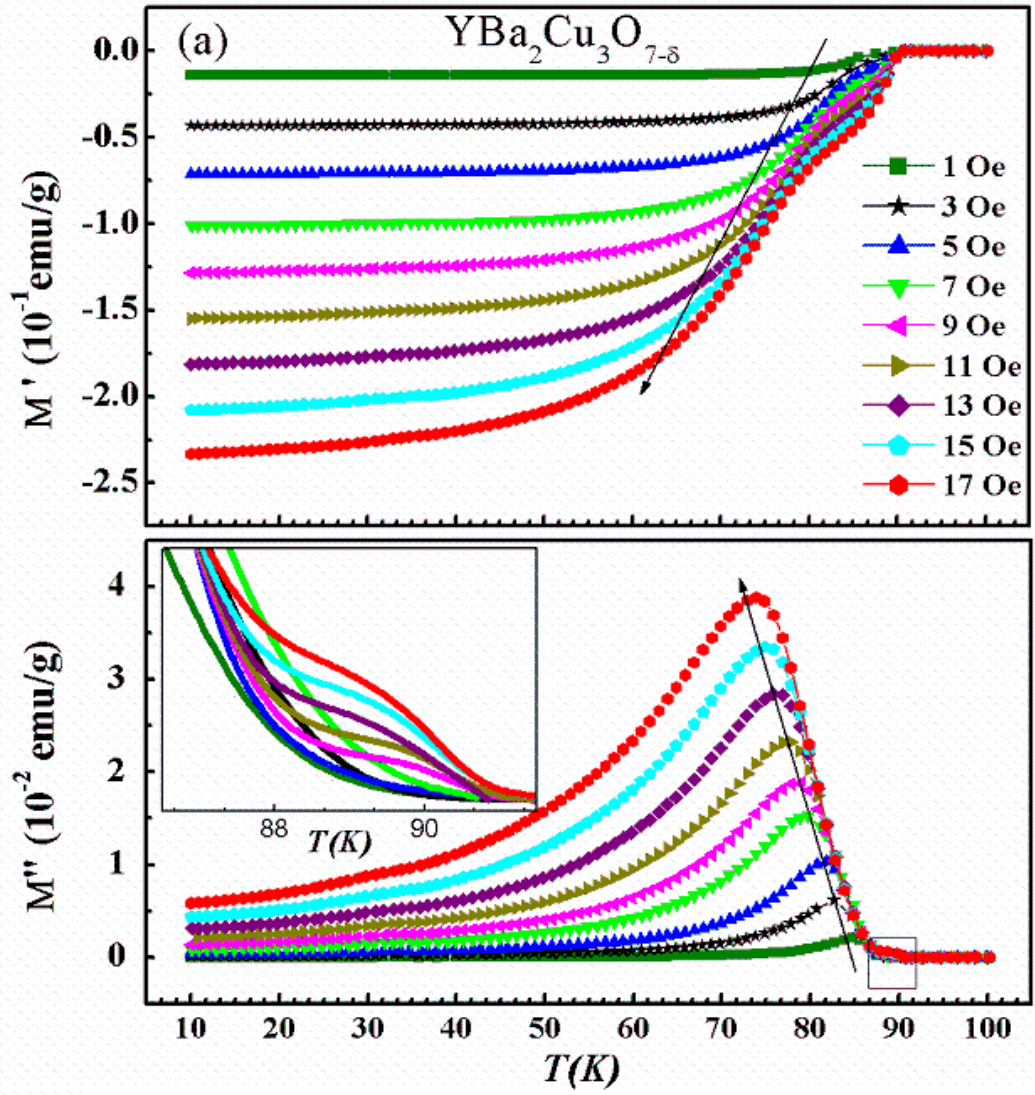

**Fig. 4(b)**

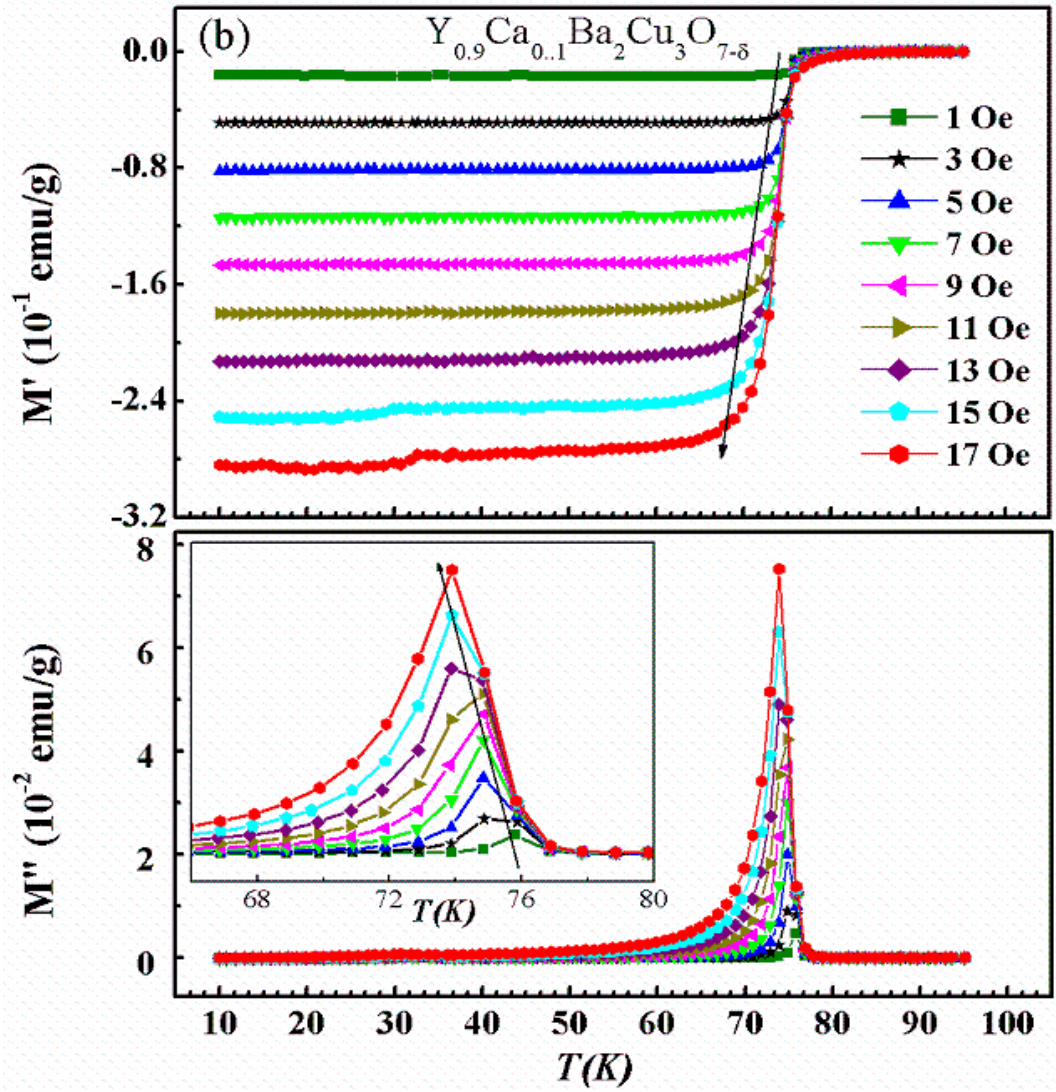

**Fig. 4(c)**

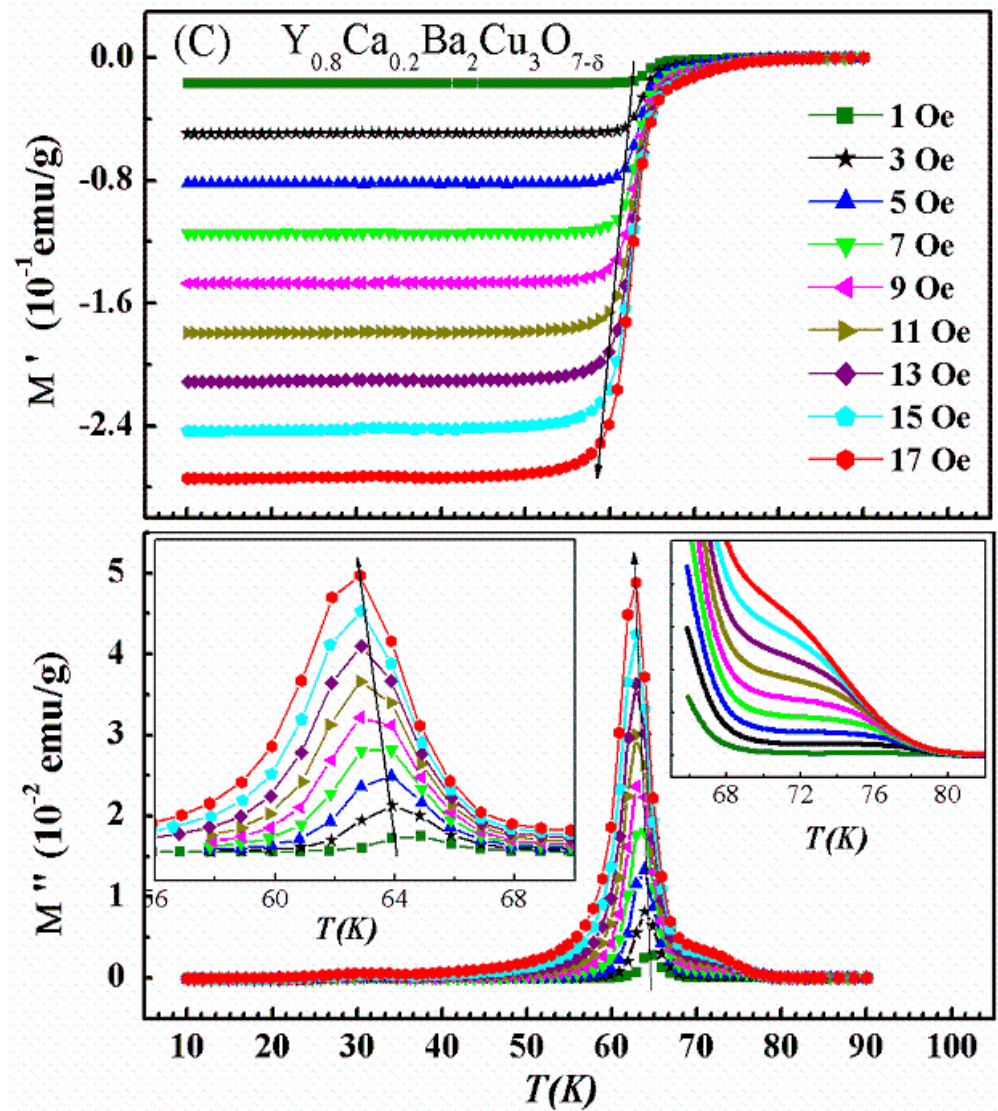

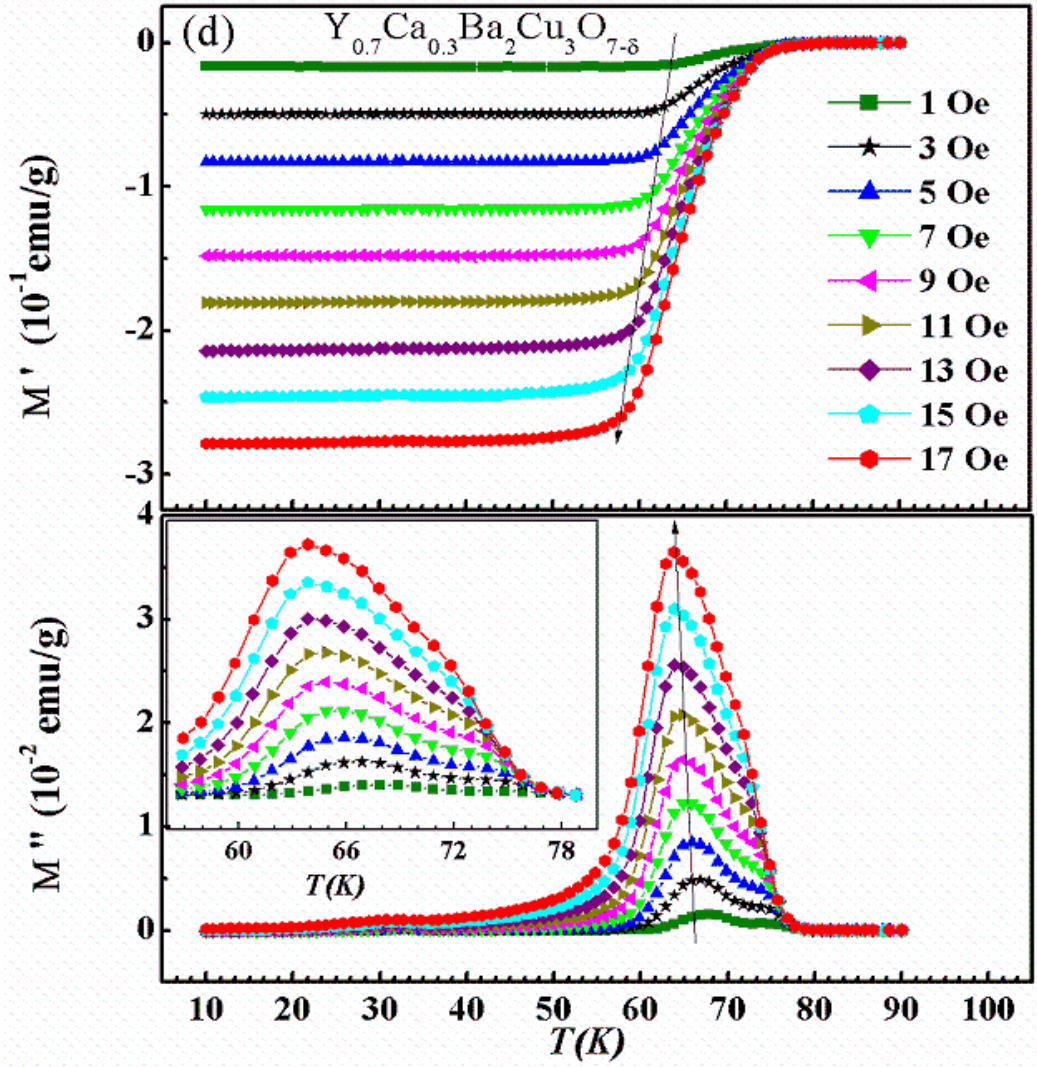

**Fig. 5(a)**

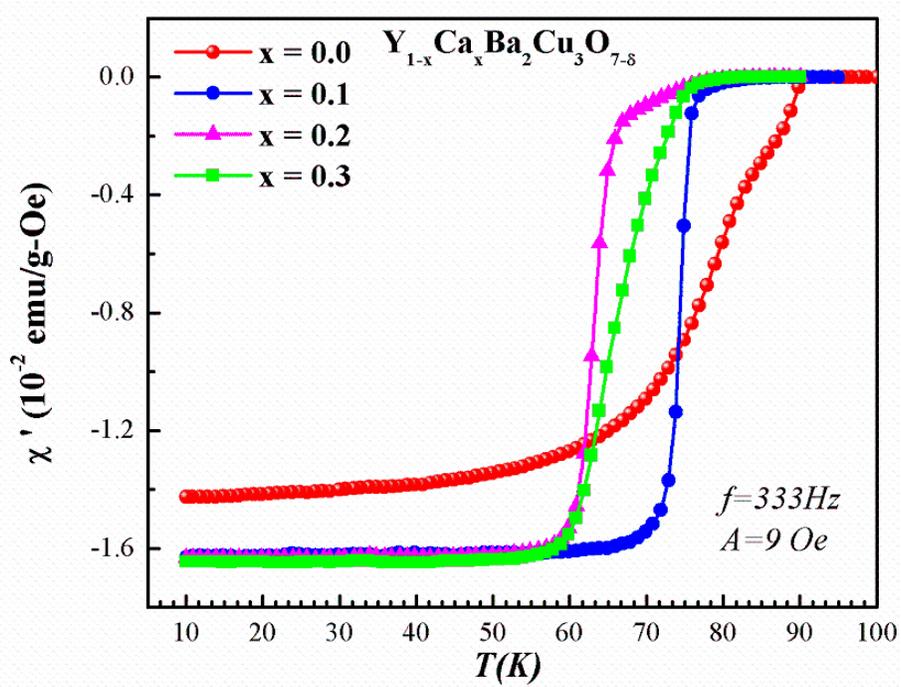

**Fig. 5(b)**

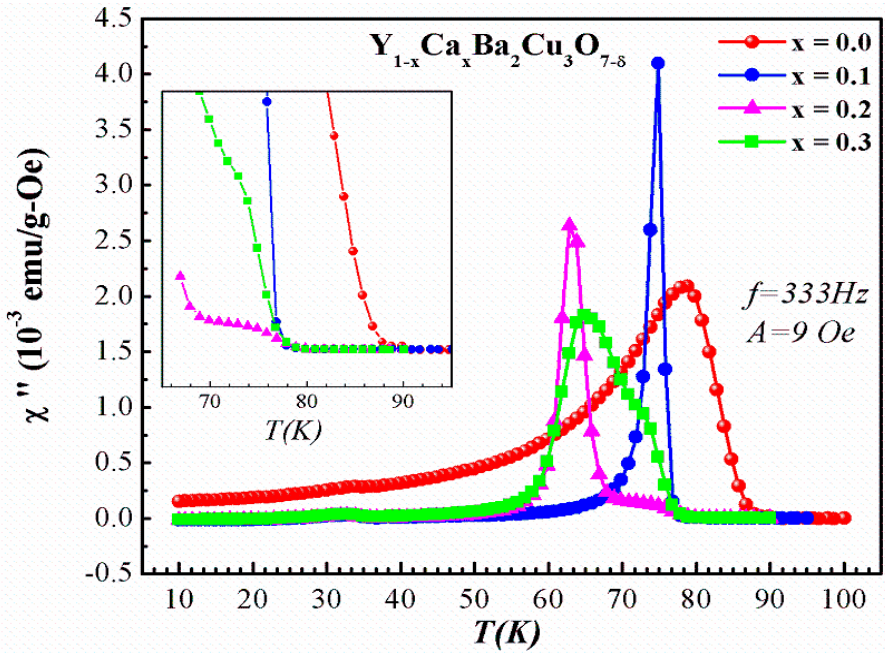

**Fig. 6**

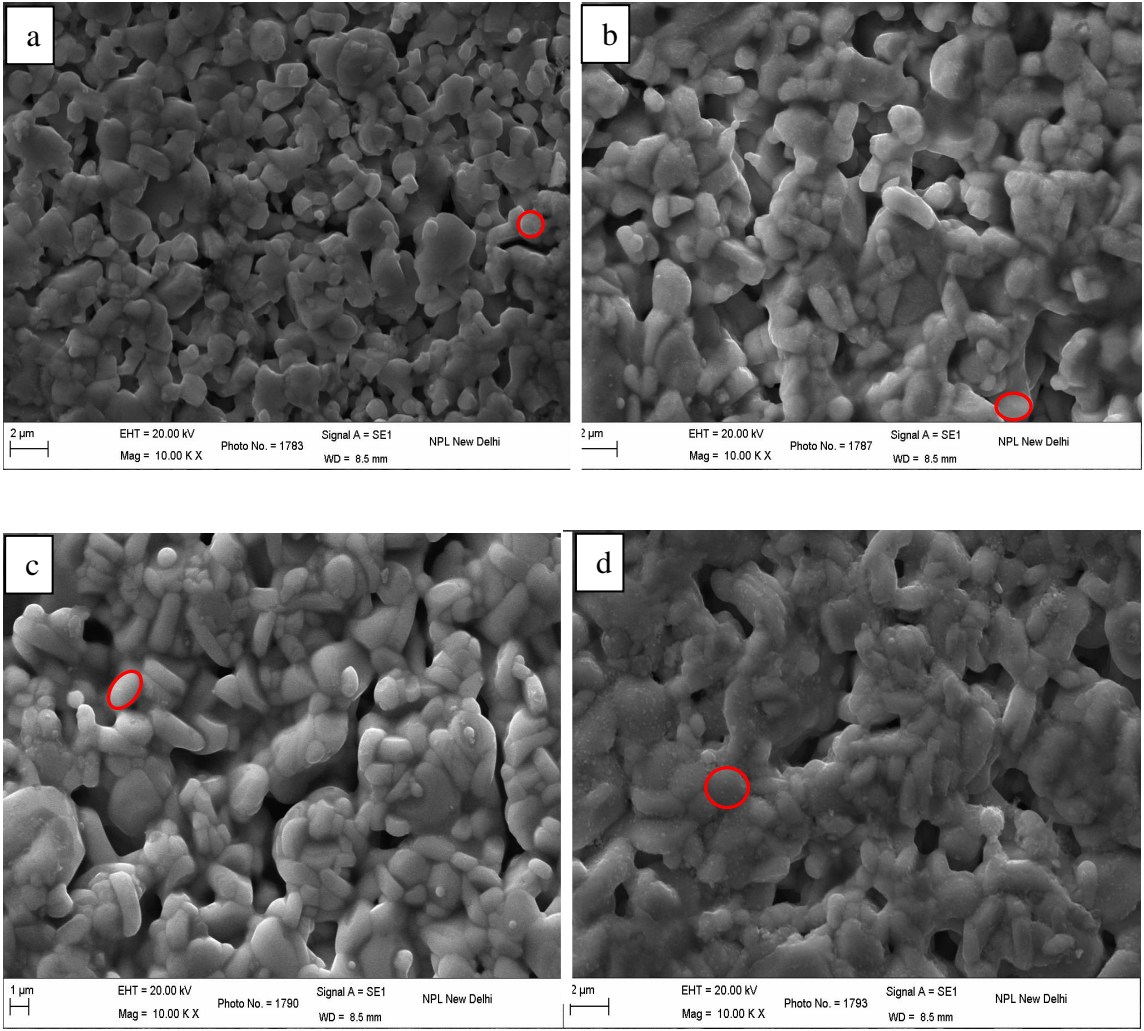